\documentclass[12pt]{article}

\newcommand{\be}{\begin{equation}}
\newcommand{\ee}{\end{equation}}
\newcommand{\bea}{\begin{eqnarray}}
\newcommand{\eea}{\end{eqnarray}}

\def\id{\protect{{1 \kern-.28em {\rm l}}}}

\def\wc{{W\Big|_{\it crit}}}

\begin{document}

\begin{titlepage}
\begin{flushright}
UCLA-02-TEP-28\\
\end{flushright}

\begin{center}
{\Large $ $ \\ $ $ \\
Exact superpotentials in $N=1$ theories with flavor 
and their matrix model formulation}\\
\bigskip\bigskip\bigskip
\bigskip

{\large ~~Iosif Bena~~~~~~~~~~~~~~~~~~~~~~~~~~~~~~Radu Roiban}

\bigskip
Department of Physics,~~~~~~~~~~~~~~~~~~~~~~~Department of Physics,\\
University of California,~~~~~~~~~~~~~~~~~~~~~~University of California, \\
~~Los Angeles,  CA 90095~~~~~~~~~~~~~~~~~~~~~~~Santa Barbara, CA 93106\\
\vskip .3cm
~~~~iosif@physics.ucla.edu ~~~~~~~~~~~~~~~~~~radu@vulcan.physics.ucsb.edu
\end{center}
\vskip 1cm
\bigskip
\bigskip\bigskip\bigskip\bigskip

\begin{abstract}
In this note we investigate the effective superpotential of an $N=1$ $SU(N_c)$ 
gauge theory with one adjoint chiral multiplet and $N_f$ fundamental chiral multiplets. 
We propose a matrix model prescription in which only matrix model diagrams with less than 
two boundaries contribute to the gauge theory effective superpotential.
This prescription reproduces exactly the known gauge theory physics for all $N_f$ and $N_c$.
For $N_f\le N_c-1$ this is given by the  Affleck-Dine-Seiberg superpotential.
 For $N_f\ge N_c+1$ we present arguments leading to the conclusion 
that the dynamics of these theories is also reproduced by the matrix model.

\end{abstract}

\end{titlepage}

\noindent
{\bf \large Introduction:} In a recent series of papers \cite{dv1,dv2,dv3}, 
Dijkgraaf and Vafa have 
proposed a perturbative method for computing the effective glueball 
superpotential of several classes of $N=1$ theories. This superpotential is 
essentially computed by summing over all the planar zero momentum Feynman 
diagrams of the theory. To better organize this computation in the case of 
fields in the adjoint representation of the gauge group, it is useful to express 
it as a matrix path integral with the 
potential given by the tree level superpotential of the original theory.

Although this duality was first obtained via a ``string theory route'' (building 
on previous work in \cite{gopa,top,cha}), it is purely a field theoretic 
duality, and very recently it has been derived within a gauge theoretic framework 
 for $SU(N_c)$ theories with adjoints \cite{grisaru}. Other related work on this 
duality has appeared in \cite{Dorey,Aganagic,Ferrari,Fuji,Berenstein,DGKV,Gorski}.

One of the most natural extensions of this duality is to theories with 
fields in the fundamental representation of the gauge group. The theory
under consideration is supersymmetric QCD with gauge group $SU(N_c)$ and $N_f$ flavors 
coupled with a single chiral field in the
adjoint representation.

One way to proceed is to represent the $N_f$ fields in the fundamental representation 
as  a $M_c\times M_f$
matrix while keeping $M_f/M_c=N_f/N_c$, and extend the matrix integral to contain 
such objects. Perturbing 
the resulting potential by mass terms for these fields allows one to integrate 
them out first and obtain an effective superpotential for the remaining 
adjoint fields. This method has been proposed in \cite{mcgr}, and has 
yet to be tested against gauge theory predictions.

Another method \cite{acfh} is based on the same integral, but uses the 
well-known fact that a quark loop in a large $N_c$ graph corresponds to adding a 
boundary to the corresponding Riemann surface. If one has $N_f$ flavors, graphs with multiple boundaries are weighted by powers of  ${N_f \over N_c} $. It is therefore 
possible to argue that, in the limit of small $N_f/N_c$, the matrix integral is 
dominated by two terms, one without quark 
loops, which is the typical Veneziano-Yankielowicz term and is proportional to the 
rank $N_c$ of the gauge group, and the other coming from the diagrams with a single 
boundary. In this limit the contributions of multiple boundaries 
are suppressed. However, for $N_f/N_c$ of order unity it is not {\it a priori} 
clear in this prescription that the diagrams with a higher number of boundaries are negligible.

This method was successfully used to compute in the matrix theory the 
effective superpotential of an $SU(2)$ gauge theory with one flavor \cite{acfh}. 
Very recently this analysis was extended to include theories with gauge group 
$SU(N_c)$ and one flavor \cite{SUZU}.
The superpotential evaluated at one of its extrema  was presented 
as a series in the bare Yukawa coupling and was shown to agree
up to 7'th order in the expansion parameter with the one 
obtained in the gauge theory after integrating out the adjoint field, adding the 
Affleck-Dine-Seiberg (ADS) nonperturbative contribution and integrating out the 
quark fields.

The puzzle that triggered our analysis was the discrepancy between the 
apparent possibility of considering a larger number of boundaries in the matrix 
model analysis and the impressive agreement between gauge theory and the 1-boundary 
superpotential. Indeed, in the analysis of \cite{acfh} diagrams with a higher number of 
boundaries are suppressed only by powers of $N_f/N_c=1/2$. Since the gauge 
theory result is exact, and the agreement is exact up to factors of the order of 
$1/2^{n}$ where $n$ is a large number, one cannot but suspect 
that this too good an agreement is due to the fact that all matrix model diagrams with 
more than a single boundary do not contribute at all.

We are therefore led 
to formulate a prescription in which only diagrams with less than two 
boundaries are relevant. This is accomplished by representing the gauge adjoint fields as
$M \times M$ matrices, and the $N_f$ fundamental fields as $M \times N_f$ natrices, and taking 
the large $M$ limit. In the case when no fields in the fundamental representation are 
present, this reduces to the original Dijkgraaf-Vafa proposal. Our proposal is very similar 
in spirit to the one in \cite{acfh}; the crucial difference is the automatic suppression of contributions of 
matrix diagrams with more than two boundaries.

The main puzzle raised by our prescription is that the matrix model does not distinguish between
$N_f<N_c$ and $N_f>N_c$ while the gauge theory physics changes drastically.

In this paper we first show that the effective superpotential obtained in the gauge 
theory by integrating out the adjoint field, adding the ADS superpotential, 
further perturbing by quark mass terms, and then integrating out the quarks 
{\it is in exact functional agreement}  with 
the effective superpotential given by the matrix model graphs 
with less than two boundaries, after integrating out the glueball superfield $S$. 
Thus, the gauge theory superpotential (with the ADS component included) 
and the superpotential obtained in the matrix model are obtained from the 
same effective superpotential $\wc(\Lambda)$ by integrating in different 
fields.

We then argue that, for $N_f>N_c$, the gauge theory superpotential evaluated at its 
extrema is also reproduced by the matrix model.

The motivation behind our proposal was the impressive agreement between gauge theory 
and the matrix superpotential due to diagrams with less than two boundaries. However, 
our computations do not exclude the (unlikely) possibility that  matrix 
diagrams with more than two boundaries  are absent only in the particular case we consider.
We will comment on alternatives at the end of our paper.

$~$

\noindent
{\bf\large Gauge theory:} It was shown by Seiberg a long time ago that the requirement of holomorphy
as well as the unbroken symmetries can be powerful allies for finding exact results
regarding effective superpotentials in $N=1$ supersymmetric theories. The simplest 
example of this sort is supersymmetric QCD. This theory contains an $SU(N_c)$
vector multiplet as well as $N_f$ pairs of quark fields, ${\tilde Q}, \,Q$,  with
${\tilde Q}$ in the antifundamental of $SU(N_c) $ and the fundamental of the $SU(N_f)$ flavor group,
 and $Q$ in the fundamental of $SU(N_c) $ and the antifundamental of a different $SU(N_f)$ flavor group.

The theory has no tree-level superpotential; the Lagrangian is thus:
\be
{\cal L}=\int d^4\theta Tr[{\bar Q}e^VQ+{\bar{\tilde Q}}e^{-V}{\tilde Q}  ]
+ \int d^2 \theta Tr[W^\alpha W_\alpha]
\ee
where the first and second traces are in flavor space while the last is in color space. 

There are several different ways of assigning charges to the various fields and coupling 
constants present in this theory. One of them, which follows by requiring that the anomaly 
cancellation condition is satisfied by the physical fields alone,
yields the following representations for the various fields, coupling constants and dynamical 
generated scale:
\be
\begin{array}{cccccccccc}
& SU(N_f) & \times & \widetilde{SU({N}_f)} & \times & U(1)_i & \times & {\widetilde{U(1)}}_i & \times & U(1)_R \\ 
Q:                 & N_f       &  &     1 &  & \;\; 1 &  & \;\; 0 & & {N_f-N_c\over N_f} \\
\tilde{Q}:         &  1 & &\bar{N}_f & & 0 &  & \;\; 1     &  & {N_f-N_c\over N_f}\\
\Lambda^{3N_c-N_f}:& 1   &&     1 &  & \;\; 1 &  & \;\; 1& & 0
\end{array}
\ee

Standard nonrenormalization theorems imply that there is no superpotential 
generated perturbatively. Nonperturbatively however, nonrenormalization theorems
fail and a superpotential is generated \cite{INSE,ADS}. Its form, 
up to numerical coefficients, is fixed 
completely by symmetries. For $N_f<N_c$ this can be argued in 3 steps.

1) First, one notices that the moduli space can be described by the gauge invariant 
meson fields $X_i{}^j={\tilde Q}_i{}^aQ_a{}^j$. Requiring invariance under both 
flavor symmetry groups implies that the effective superpotential can be a function 
only of $\det X$.

2) The requirement that the $R$-charge of the superpotential is $2$ implies that 
the exponent of $\det X$ is ${1\over N_f-N_c}$.
\vspace*{-.3cm}

3) Finally, on dimensional grounds, one has to add $\Lambda^{3N_c-N_f\over N_c-N_f}$,
where $\Lambda$ is the dynamically generated scale of the theory.

The fact that the dynamical scale of the theory does not appear under a 
logarithm implies indeed that this superpotential is nonperturbatively generated.
The coefficient of the resulting term can be explicitly computed 
for $N_c-N_f=1$ (\cite{ADS} for the $SU(2)$ gauge group and \cite{Cordes} for $SU(N_c)$) 
and is found to be equal to unity. For $N_c-N_f>1$ the coefficient can be 
obtained by adding mass terms to the appropriate number of flavors and
integrating them out. The result is the usual ADS superpotential \cite{INSE}:
\be
W_{dyn}=(N_c-N_f) \left({\Lambda^{3N_c-N_f}\over \det X}\right)^{1\over N_c-N_f}
\ee

A slight deformation of the theory we are discussing is the inclusion of a tree 
level superpotential:
\be
{1\over 2}a Tr[{\tilde Q} Q{\tilde Q}Q] = {1\over 2}a Tr[X^2]
\ee
For consistency with the various symmetries of the original theory, the charge assignments for 
this new coupling constant are:
\be
\begin{array}{cccccccccc}
& SU(N_f) & \times & \widetilde{SU({N}_f)} & \times & U(1)_i & \times & {\widetilde{U(1)}}_i & \times & U(1)_R \\ 
a:         & 1      &  &     1 &  & \;\; -2 &  & \;\; -2 & & 2{N_f-N_c\over N_f} 
\end{array}
\ee
This superpotential can be thought of as arising from integrating out a massive adjoint field with
a cubic coupling with the quarks.

One can show that in this theory the only superpotential that can be generated is again the ADS 
one. Furthermore, adding mass terms to some of the quarks and integrating them out 
produces a new effective action which has the same form as the original one. This is achieved 
by keeping the coupling constant $a$ fixed while taking the quark mass to infinity and keeping a 
certain combination of this mass and dynamically generated scale fixed. This procedure also yields
the change in the dynamical scale to be:
\be
({\Lambda'}){}^{3N_c-N_f'}={\Lambda}{}^{3N_c-N_f}\det m
\label{matching}
\ee
where $m$ is the mass of the quarks which were integrated out.

The central object $\wc$ that contains all the information about the dynamics of the theory can be obtained
by integrating out all the fields. The resulting function depends on the 
scale $\Lambda_0$ \footnote{The index denotes the fact that there are no more light fields 
in the theory.} as well as on the coupling constant $a$ and the masses of the quark 
fields\footnote{If one interprets the quartic tree level superpotential as arising from integrating 
out an adjoint field, then the coupling constant $a$ is a function of the mass of the adjoint field 
and the strength $g$ of the trilinear coupling $Tr[{\tilde Q}\phi Q]~$; $~a={g^2\over M}$.}.
Knowledge of this function allows one to reconstruct the full superpotential via Legendre transforms
\cite{INSE}.

Without loss of generality we can integrate out all the quark fields at the same time, by introducing a 
mass matrix $m$ proportional to the identity matrix. It is certainly possible to introduce hierarchical 
masses (and/or nondiagonal mass matrices) and integrate out one quark field at a time. However, the final 
result will be expressed only in terms of the scale $\Lambda_0$; the expression in terms of 
more general mass matrices can be trivially restored.

It is customary to write the superpotential in terms of the meson field $X={\tilde Q}Q$. Then, 
the superpotential deformed by mass terms is:
\be
W_{eff}= mTr[X] - {a\over 2} Tr[X^2] + (N_c-N_f)\left[{\Lambda_{N_f}^{3N_c-N_f}
\over \det X} \right]^{1\over N_c-N_f}
\ee
where $\Lambda_{N_f}$ is the scale of the theory with $N_f$ light fields. 
Since we assumed that the mass matrix is proportional to the 
identity matrix, it follows that, at the minimum of $W_{eff}$, $X$ will have the same 
property. Thus, we can write from the outset 
\be
X=x \id_{N_f}~~.
\ee

By simple inspection of the superpotential above it is clear that explicitly 
finding its extrema for general numbers of colors and flavors is not an easy 
operation. The approach we take is to use 
the equation of motion for $x$ to cast the superpotential into a simple form in 
which one still has to replace $x$ by the solution of some algebraic equation. 

The equation for $x$ is simply:
\be
m - a x -
 {\Lambda_{N_f}^{3N_c-N_f\over N_c-N_f}
\over x^{N_c\over N_c-N_f}}=0~~.
\label{origeom}
\ee
It is convenient to introduce new dimensionless variables
\be
y=\left[{m\over \Lambda_{N_f}^{2k+1}}\right]^{1\over k} x~,~~~~
\beta={a\over m}\,\left[{\Lambda_{N_f}^{2k+1}\over m}\right]^{1\over k}
\equiv {g^2\over m^2M}\Lambda_0^{3}~,~~~~k\equiv {N_c\over N_c-N_f}~,
\ee
where $\Lambda_0$ is the dynamical scale of the theory with no light fields\footnote{
Indeed, by taking $N_f'=0$ in equation (\ref{matching}) and using the definition of $k$
it is easy to see that
\be
\Lambda_0^{3}=m\left[{\Lambda_{N_f}^{2k+1}\over m}\right]^{1\over k} \nonumber
\ee}.
In terms of these variables the equation of motion for $x$ becomes:
\be
\beta y^{k+1}-y^k + 1=0~~,
\label{c1}
\ee
or equivalently
\be
y^{-k}=1-\beta y~~,
\label{constrG}
\ee
since $y=0$ is not a solution of equation (\ref{c1}).

Now we use these equations to eliminate the quadratic term in $W_{eff}$.
A small amount of algebra leads to the superpotential evaluated at its minimum 
in terms of the newly introduced variables $y$ and $k$:
\be
\wc={1\over 2} N_f\Lambda_0^3\left[y+{k+1\over k-1}{1\over  y^{k-1}}
\right]
\label{finalG}
\ee
where again $y$ is a solution of equation (\ref{constrG}).

This is the form that will be compared with the matrix model predictions. In the next section,
using our modified extension of the Dijkgraaf-Vafa prescription to include fields in the fundamental 
representation, we will compute the value of the superpotential at its critical points and recover 
equations (\ref{finalG}) and (\ref{constrG}) {\it for all $N_c$ and $N_f$}.
We will then present a gauge theoretic argument that (\ref{finalG}) and (\ref{constrG}) are 
also true for $N_f > N_c$. \footnote{The case $N_c=N_f$ is problematic 
since the ADS superpotential does not admit a continuation to this point.}

$~ $

\noindent
{\bf\large The matrix model:} 
The prescription of Dijkgraaf and Vafa instructs that to compute the effective 
superpotential for the gaugino condensate, we have to compute the planar partition function for the 
matrix model with a potential which is the tree level superpotential of the $N=1$ theory we
are interested in. The original arguments covered theories which had fields in bi-fundamental 
representations of the gauge group. 

Applied to a superpotential with a single critical point, this proposal yields
the  effective superpotential for the glueball superfield $S$ by the following three steps:

First, one computes the contribution to the free energy due to planar diagrams, ${\cal F}_0$. 
This is accomplished
by formally replacing the gauge theory fields with $M\times M$ matrices in the matrix model potential.
Then one computes the path integral in the limit $M\rightarrow\infty$. 

The second step is to identify the 't~Hooft coupling in the matrix model with the gauge theory 
glueball superfield. At this stage ${\cal F}_0$ becomes a function of $S$ only.

The third and last step if to construct the gauge theory effective superpotential as:
\be
W_{DV}=N_c  {\partial {\cal F}_0\over \partial S} + \tau S~~. 
\ee
The only gauge theory ingredients entering this relation are $N_c$ - the number of colors, 
and $\tau$ - the bare coupling constant. There is no relation between $N_c$ and the 
dimension of the matrices used in the matrix model computations; the first one is a parameter while the
second one was identified with the glueball superfield.

In the case of a trivial superpotential the only contribution to ${\cal F}_0$ comes from the 
normalization of the matrix path integral by the volume of the gauge group. In this case 
$W_{DV} $ reproduces the Veneziano-Yankielowicz superpotential.

Attempts to extend this prescription to include  fields in the fundamental representation of the
gauge group were formulated in \cite{mcgr}, \cite{acfh}.  The idea is again to use the tree level 
superpotential as potential for the matrix model. In one proposal 
both the number of flavors $N_f$ and the number of colors $N_c$  are ``promoted'' to  matrix 
model variables  $M_f$ and  $M_c$ such that $N_f/N_c=M_f/M_c$ \cite{mcgr}. In the other proposal 
none of them is \cite{acfh}. In both cases, in the limit $N_f \sim N_c$ one cannot perform an expansion
 in the number of boundaries in the  matrix model, which makes comparison with gauge theory difficult.

Here we propose that the matrix model variables are $M \times M$ matrices if the corresponding 
gauge theory fields are in the adjoint representation and $M \times N_f$ matrices if there are $N_f$ fields 
in the fundamental representation in the gauge theory. Then, as in the original DV prescription, 
we identify the gauge theory glueball superfield with the  't~Hooft coupling of the matrix model, i.e.
$S=g^2 M$. As put forward in \cite{acfh}, the extended DV superpotential is the sum of the original 
one and extra contributions coming from matrix diagrams with boundaries:
\be
W_{DV}=N_c  {\partial {\cal F}_0\over \partial S} + \tau S + {\cal F}_{boundaries}
\label{proposal}
\ee

If one recalls that $S$ was identified with the dimension of the matrices and examines the $M$ 
dependence of the various terms in the equation (\ref{proposal}), it becomes clear that {\it only}
diagrams with one boundary contribute in the planar (large $M$) limit. Indeed, equation (\ref{proposal})
can be organized as an expansion in $(N_f/M)$. Then, due to the derivative with respect to $S$, the first 
term and the contribution of diagrams with one boundary will be of the same order in an $1/M$ expansion.
{\it All diagrams with two or more boundaries are suppressed by 
factors of $1/M$ and do not contribute in the large $M$ limit.}

As in the original DV proposal, the part of ${\cal F}_0$ arising from the volume of the gauge group
yields the Veneziano-Yankielowicz superpotential. This part is not modified by the inclusion of fields 
transforming in representations of flavor symmetry groups because these symmetries are only global.

For supersymmetric QCD with an adjoint field $\phi$ this leads to \cite{mcgr}, \cite{acfh}:
\bea
e^{\cal F}&=&\int D\phi DQD{\tilde Q} e^{-\left[{1\over 2} M_\phi Tr[\phi]^2 + Tr[m{\tilde Q}Q]
+Tr[g{\tilde Q}\phi Q]\right]}\nonumber\\
&=&
\int D\phi DQD{\tilde Q} e^{-\left[Tr[m{\tilde Q}Q]
-{1\over 2}Tr[a{\tilde Q} Q{\tilde Q} Q]\right]}
\eea
with $a={g^2\over M_\phi}$, as in the gauge theory discussion.
The path integral above can be computed in both of its forms, using the analysis 
of  \cite{french}.
Using the first form above and the analysis in \cite{acfh} we can show that  
the contribution to the free energy of the matrix model in the large $M$ limit is:
\be
{\cal F}_{\chi=1}=-N_f S\left[{1\over 2}+{1\over 4\alpha S}(\sqrt{1-4\alpha S}-1)
-\ln\left[{1\over 2}+{1\over 2}\sqrt{1-4\alpha S}\right]\right]
\ee
where $\alpha={a\over m^2}$.
Then, according to our extended DV prescription, the superpotential is:
\bea
W_{DV}\!\!&=&\!\!
N_cS\left[1-\ln {S\over \Lambda^3}\right]+{\cal F}_{\chi=1}=\label{DVsuper}\\
&&\!\!\!\!\!\!\!\!\!\!\!\!\!\!\!\!\!\!\!\!\!\!\!\!\!\!\!\!
=N_c S(1-\ln{S\over\Lambda^3})
-N_fS\left[{1\over 2}+
{1\over 4\alpha S}(\sqrt{1-4\alpha S}-1)-\ln{1\over 2}(1+\sqrt{1-4\alpha S})\right]~,
\nonumber
\eea
for all $N_f$ and $N_c$.

As in the gauge theory case, we proceed by integrating out the massive fields. Various terms in 
the equation above imply that the gaugino fields acquired a mass and thus can be integrated out.
Despite the complicated form of the superpotential, its extrema are given by a very simple equation:
\be
N_c\ln {S\over \Lambda^3}=N_f\ln {1\over 2}(1+\sqrt{1-4\alpha S})~~.
\ee
The first step is to use this equation to eliminate the logarithms in $W_{eff}$.
Then, the critical point equation can be cast in a more useful form:
\be
\sqrt{1-4(\alpha S)} =
2\left({(\alpha S)\over \alpha\Lambda^3}\right)^{k\over k-1}-1
\label{constrM}
\ee
which in turn leads to the following expression for the superpotential at the critical points:
\bea
\wc={N_f \over 2\alpha}\left[ {k+1\over k-1} (\alpha S) -
\left({(\alpha S)\over \alpha\Lambda^3}\right)^{k\over k-1}
+1\right]
\eea

This equation does not appear similar to the corresponding one on the gauge theory side. It is
nevertheless possible to relate them more closely. To this end we massage equation (\ref{constrM}). 
A small amount of algebra combined with the observation that $S=0$ is not a solution of 
equation (\ref{constrM}),
casts it in the following form:
\be
1
-\left({(\alpha S)\over \alpha\Lambda^3}\right)^{k\over k-1}=
(\alpha \Lambda^3)\left({(\alpha S)\over \alpha\Lambda^3}\right)^{-1\over k-1}
\label{modconstr}
\ee
Introducing the notation
\be
w= \left({(\alpha S)\over \alpha\Lambda^3}\right)^{-1\over k-1} ~~,
\ee
equation (\ref{modconstr}) becomes:
\be
w^{-k}=
1- (\alpha \Lambda^3)w
\label{nume}
\ee
which is the same as equation (\ref{constrG}).

In terms of this new variable the superpotential evaluated at its minimum
is given by:
\be
\wc={1 \over 2}{N_f \Lambda^3}\left[ {k+1\over k-1} {1\over w^{k-1}} 
+w
\right]
\label{finalM}
\ee
where $w$ must be replaced by a solution of equation (\ref{nume}).

Since $w$ in the matrix model result as well as $y$ on the gauge theory side are 
dummy variables, the results of the two computations agree provided that the matrix model 
scale $\Lambda$ is identified with the gauge theory scale in the confining vacua as:
\be
\Lambda^3=\Lambda_0^3=(\det m)^{1\over N_c}{\Lambda_{N_f}^{1+2k\over k}}~~,
\label{finalmatching}
\ee
which is consistent with both theories not having massless fields. Furthermore, since the analysis 
above goes through if $N_f=N_c$ this seems to suggest that the same should hold in the gauge 
theory as well.

${}$

\noindent
{\bf\large ${\bf N_f >N_c}$:} 
It is clear from equation (\ref{DVsuper}), and it was also pointed out in \cite{acfh} that 
the extended DV superpotential is insensitive to the range of $N_c$ and $N_f$.  It was suggested 
in \cite{acfh} that this ``problem'' might be solved by including contributions from 
additional boundaries, which in the $N_f \rightarrow N_c$ limit would become important.
The proposal we put forward in this paper implies that such contributions are inexistent.

We have shown above that the 1-boundary matrix integral reproduces the expected gauge 
theory answer for {\it all} $N_f$ and $N_c$ for which the ADS superpotential holds. 
Moreover, this range can be extended to include also the case $N_f=N_c+1$. 
Indeed, without introducing baryonic sources, the superpotential in this case is just the 
continuation of the ADS one \cite{SEI1} to this number of flavors.

The $\wc$ obtained in the matrix model always looks as if one had blindly started in gauge theory from an 
ADS-like superpotential, even for $N_f>N_c$. Our proposal implies that, even though 
for $N_f>N_c+1$ no superpotential is generated in the beginning,
once all quark fields are integrated out the superpotential evaluated at the critical point is still
given by equations (\ref{finalM}) and (\ref{nume}).
At first glance this might seem problematic. It is nevertheless possible to argue that there 
is no contradiction with the field theory analysis \cite{SEI2}.

The argument is a slight generalization of the 
analysis in \cite{SEI2} and is based on Seiberg's duality together with the possibility of 
freely passing from electric variables to magnetic variables in the 
path integral.

In gauge theory, one can deform the theory by adding mass terms for all quark fields 
and then integrate them out. For $N_f\ge N_c+2$ the gauge theory is strongly coupled and 
the correct description is given by its Seiberg dual. In this description we could 
add masses to the magnetic quarks and integrate them out, until we reach the theory with
a completely broken gauge group. The superpotential for this theory is written in terms of 
the fields which are dual to the electric meson fields and, in the absence of baryonic sources, 
is just the ADS superpotential. At this point we dualize back to the electric theory which is 
now weakly coupled.

The results of the gauge theory analysis for $N_f = N_c+1$, given by  (\ref{finalG}) and (\ref{constrG}),
can now  be applied without reservations. Since the matching of scales when fields are integrated 
out is fixed by the renormalization group equations,
it follows that the effective scale when all quarks are integrated out is given by
equation (\ref{matching}) with the initial $N_f$ and $N_c$ \cite{SEI1,SEI2}. 

Thus, by integrating out all fields in the fundamental representation, we obtain 
the same $\wc$ as if we blindly started with an electric ADS-like superpotential for 
$N_f\ge N_c+2$ and integrated out all quark fields. 
This explains the apparently unnatural agreement between the matrix model results 
and the introduction of an ADS-like superpotential for $N_f\ge N_c+2$.

${}$

\noindent
{\bf\large Discussion:} In this paper we formulated an extension of the Dijkgraaf-Vafa 
proposal which includes fields in the fundamental representation. Basically, this proposal 
states that {\em only matrix model diagrams with 
less that two boundaries contribute to the 
``gauge theory - matrix model'' duality}. We have tested this proposal by analyzing in detail 
supersymmetric QCD  coupled to an adjoint chiral multiplet, both from the gauge theory
and the matrix model perspective. We have found that, in the range of parameters where 
the Affleck-Dine-Seiberg superpotential is valid, it is reproduced exactly by the matrix
model computation. Using Seiberg's duality we have then argued that this agreement persists
even for $N_f>N_c$, when the gauge theory is strongly coupled.
Therefore the gauge theory and the matrix model analysis described earlier in 
this note agree for all values of $N_c$ and $N_f$. 

Since for theories with
$N_f\ge N_c$ the baryons acquire a nonvanishing expectation value, it would be interesting 
to repeat the analysis by including baryonic sources both in the matrix model and in 
the gauge theory. This would extend the number of arguments of $\wc$
and would  provide a further test of the extension of the 
DV prescription proposed in this note.

It would also be interesting to see how our proposal would be implemented when the recent field-theoretic 
proof \cite{grisaru} of the initial DV proposal is extended to include fields in the 
fundamental representation. Another worthwhile endeavor would be finding the gauge theory 
interpretation of matrix model diagrams 
with more than one boundary. According to our proposal they are suppressed 
like the nonplanar diagrams in the original 
matrix model;  it seems however likely that they do not correspond to gravitational corrections, given the 
difficulties with the geometric engineering of such theories.

As we pointed out in the Introduction, our computations do not exclude the (unlikely) 
possibility that  matrix 
diagrams with more than two boundaries  are only absent in the particular case we consider.
We can imagine several possible reasons for this:

1) The extension of the DV proposal to include fundamental fields is still such that only diagrams
with a single boundary contribute. Nevertheless this could be due to multiple 
boundaries being considered as disconnected diagrams, and thus
not contributing to the free energy.

2) Diagrams with more than one boundary simply vanish. If this were the case, this would give a
highly nontrivial prediction for the matrix model.

Whatever the cause, we believe that the impressive agreement between the gauge theory and matrix model 
results strongly supports the fact that matrix model diagrams with more than one boundary do not 
contribute to the duality, and we look forward to seeing how this phenomenon will emerge within a purely 
gauge theoretic framework.

${}$

\noindent
{\bf Acknowledgments:}
We would like to thank David Gross, Per Kraus, Joe Polchinski, Johannes Walcher and 
Sebastian de Haro, for useful comments and discussions.
The work of I.B. was supported by the NSF under 
Grant No. PHY00-99590. The work of R.R. was supported in part by DOE under 
Grant No. 91ER40618 and in part by the NSF under Grant No. PHY00-98395.

\end{document}